\documentclass[conference]{IEEEtran}
\IEEEoverridecommandlockouts

\usepackage{cite}
\usepackage{amsmath,amssymb,amsfonts}
\usepackage{graphicx}
\usepackage{textcomp}
\usepackage{xcolor}
\usepackage{graphicx}
\usepackage{multirow}
\usepackage{amsmath} 
\usepackage{booktabs} 
\usepackage{subcaption} 
\usepackage{amsmath}
\usepackage{graphicx}
\usepackage{booktabs} 
\usepackage{array}    
\usepackage{threeparttable}
\usepackage{tablefootnote}
\usepackage{algorithm}
\usepackage{algpseudocode}
\usepackage{booktabs} 
\usepackage{multirow} 
\usepackage{physics}
\usepackage{adjustbox}

\def\BibTeX{{\rm B\kern-.05em{\sc i\kern-.025em b}\kern-.08em
    T\kern-.1667em\lower.7ex\hbox{E}\kern-.125emX}}
\begin{document}


\title{Deep Adversarial Learning with Activity-Based User Discrimination Task for Human Activity Recognition \\
}

 \author{\IEEEauthorblockN{Francisco M. Calatrava-Nicol\'as\IEEEauthorrefmark{1}, Shoko Miyauchi\IEEEauthorrefmark{2},  Oscar Martinez Mozos\IEEEauthorrefmark{1}\IEEEauthorrefmark{3}}
 \IEEEauthorblockA{\IEEEauthorrefmark{1}AASS, Örebro University, 70182 Örebro, Sweden\\
 Email: \{francisco.calatrava-nicolas, oscar.mozos\}@oru.se}
  \IEEEauthorblockA{\IEEEauthorrefmark{2}Faculty of Information Science and Electrical Engineering, Kyushu University, 819-0395 Fukuoka, Japan\\ Email: miyauchi@ait.kyushu-u.ac.jp }
 \IEEEauthorblockA{\IEEEauthorrefmark{3}ETSIDI, Universidad Politéctina de Madrid, 28012 Madrid, Spain\\ Email: oscar.mmozos@upm.es}
 \thanks{Corresponding authors: Francisco M. Calatrava-Nicol\'as (francisco.calatrava-nicolas@oru.se) and Shoko Miyauchi (miyauchi@ait.kyushu-u.ac.jp) }

}

\maketitle

\begin{abstract}
We present a new adversarial deep learning framework for the problem of human activity recognition (HAR) using inertial sensors worn by people. Our framework incorporates a novel adversarial activity-based discrimination task that 
addresses inter-person variability—i.e., the fact that different people perform the same activity in different ways. Overall, our proposed framework outperforms previous approaches on three HAR datasets using a leave-one-(person)-out 
cross-validation (LOOCV) benchmark. Additional results demonstrate that our discrimination task yields better classification results compared to previous tasks within the same adversarial framework..
\end{abstract}


\section{Introduction}
\IEEEpubidadjcol

The problem of human activity recognition (HAR) consists of identifying the activities carried out by a 
person by analyzing the data collected with different pervasive sensors \cite{chaquet2013survey,dang2020sensor}. HAR is applied to different research areas such as recognition of activities of the daily living (ADL) \cite{debes2016monitoring,pirsiavash2012detecting}, human-robot interaction \cite{piyathilaka2015human,roitberg2014human}, autonomous vehicles \cite{ohn2016looking,rasouli2019autonomous}, healthcare \cite{wang2019survey,zhou2020deep}, or surveillance \cite{chaquet2013survey,yeffet2009local}. 

This paper focuses on the application of HAR to classify ADL using data from wearable inertial sensors. Although this problem has been lately addressed using different deep learning approaches \cite{calatrava2023ieeesensors,chen2020metier,Fabio_Hernandez,ordonez2016deep,YangWang,yang2015deep}, it still presents several challenges. The first one is the lack of generalization to novel and unseen users due to the heterogeneity in data distributions arising from inter-person variability, i.e., different individuals perform the same task differently \cite{barshan2016investigating,Plotz2023Percom}. 
A potential solution to this problem would involve collecting and annotating extensive volumes of data from different people carrying daily life activities in order to mirror real-world conditions, but this would require extra effort and resources. Consequently, recent HAR research efforts focus on reducing data distribution shifts caused by inter-person variability, with the goal of enhancing classification performance without additional data collection. A secondary challenge arises directly from the design of deep learning models: as more data from new users is incorporated to improve model generalization, the complexity of these models often increases. 
Finally, models that analyse data coming directly from the sensors worn by users may learn features that highly discriminate user-information unintentionally, which increases the risk of information-disclosure \cite{iwasawa2017privacy}. 

To address the previous issues, representation learning approaches based on multi-task, adversarial learning and self-supervised learning have been proposed \cite{ShuSungho2022Percom,chen2020metier,bai2020adversarial,qin2023generalizable}. These approaches aim at increasing generalisation and, in the case of adversarial approaches, potentially mitigate concerns related to subject information leakage\cite{iwasawa2017privacy}. In particular, the system in \cite{chen2020metier} integrates activity and user recognition within a multi-task learning framework,
however, its generalization to new users is uncertain because it lacked LOOCV evaluation. In addition, this approach uses direct user information for the user recognition task which may lead to privacy issues. Finally, the addition of new participants to the training increases the classifier's complexity since it contains as many classes as people in the dataset,
which can affect scalability. Further work in \cite{bai2020adversarial} proposes an adversarial framework with a binary discriminator that determines whether a pair of random activity feature vectors originate from the same participant or from different ones. However, this random selection may hinder its generalization capabilities. Alternatively, the work in \cite{ShuSungho2022Percom} proposes an adversarial discrimination task to recognize each participant individually, which also grows in complexity with the number of participants. Finally, the approach in \cite{qin2023generalizable} employs self-supervised auxiliary tasks to enhance generalization in human activity recognition, nevertheless they apply a modified version of the LOOCV which limits direct comparisons with other methods.
\IEEEpubidadjcol

In this paper, we present an approach that provides new improvements to the previous challenges. 
To this end, we propose an adversarial framework with a novel activity-based user discrimination task that, in contrast to \cite{ShuSungho2022Percom,bai2020adversarial}, includes not only the user dimension but also the activity dimension in order to allow a better integration of the inter-person variability concept into the learned model. 
Our primary insight is that embedding the activity within the training discriminator dataset's construction logic will facilitate finding a common deep feature space for the same activity while intrinsically reducing the inter-person variability for that activity. In order to do that,  we propose a binary classification task consisting of determining whether a pair of activity feature vectors come from the {\em same} person and the {\em same} activity (class 1), or from {\em different} persons but still the {\em same} activity (class 0). Our binary classification approach does not increase the discriminator complexity with more participants as in \cite{ShuSungho2022Percom,chen2020metier}. Moreover, our proposed framework can mitigate privacy issues since, in contrast to \cite{chen2020metier}, we do not recognize specific users. In addition, and opposite to \cite{bai2020adversarial}, we propose the use of a sampling strategy 
to ensure generalization across all users. Finally, our adversarial framework incorporates the feature extractor with a reduced parameter count from \cite{calatrava2023ieeesensors}. 

We evaluate our framework in a Leave-One-(Person)-Out Cross-Validation (LOOCV) setup ensuring that our model is tested against new completely unseen individuals. LOOCV is supported by previous works \cite{hammerla2015let,calatrava2023ieeesensors} as an effective method for simulating real-world conditions. 
Experimental results show the improvement in classification results with respect to previous approaches by using our adversarial model. Moreover, we show a reduction of the inter-person variability gap for the same activity between the training and testing participants when using our proposed discriminator task.

\section{Related Work}

The HAR problem applied to the classification of ADL using wearable inertial sensors has been lately addressed using different deep learning approaches. For example, convolutional neural networks (CNN) were used in \cite{calatrava2023ieeesensors,Cruciani,yang2015deep}, in which signals are divided into windows and classified as a one-shot input. In addition, extra attention mechanisms are also included in the deep learning framework as in \cite{YangWang}. Moreover, recurrent neural networks such as long short-term memory (LSTM) \cite{Fabio_Hernandez} or Gated Recurrent Units \cite{Ullah} have been applied to take into account the temporal dynamics of the input signals. In addition, hybrid models have been applied to the HAR problem combining the advantages of the previous approaches. For example, the work in \cite{Ullah} presents a model based on stacked LSTM to capture the temporal features. Also, the work in \cite{ordonez2016deep} proposes a hybrid framework that aims to exploit and capture the temporal dynamics of the movements. Moreover, in \cite{Khatun}, a hybrid model is presented integrating CNN+LSTM with a self-attention mechanism to better capture both, spatial and temporal features.

Multi-task learning has also been applied to the HAR problem based on the premise that tackling multiple tasks simultaneously can lead to improvements in the performance of each individual task. For example, authors in \cite{peng2018aroma} propose a multi-task framework to recognize two types of activities: simple and complex. Moreover, a multi-task learning framework is introduced in \cite{chen2020metier} that integrates activity and user recognition within the learning framework to improve the classification, although its generalization to new users is uncertain because it does not provide a leave-one-person-out evaluation. Moreover, that approach applies the recognition task to data coming directly from the user which may lead to potential privacy issues. Finally, the classifier's complexity in \cite{chen2020metier} increases with the addition of new participants.

More recently, adversarial networks have been proposed to classify ADL. Adversarial models try to find a common deep feature space that generalizes well among the different users. This is typically achieved through an adversarial user-based discrimination task that results in a deep feature vector which is less user-discriminant and thus, improves the privacy of the users \cite{iwasawa2017privacy}. In this sense, the work in \cite{bai2020adversarial} proposes an adversarial binary discrimination task that determines whether two feature vectors come from the same participant or from different ones. The goal is to reach a feature space in the feature extractor that deceives the discriminator, making the features more invariant to individual users. However, this random selection method does not include all possible user combinations, which could limit its ability to generalize across all subjects.  To overcome that, the work in \cite{ShuSungho2022Percom} proposes an adversarial discrimination task in which a discriminator is trained to recognize each participant individually getting state-of-the-art results. In this case, however, the classifier also grows in complexity with the addition of new participants. 

Finally, the approach in \cite{qin2023generalizable} employs self-supervised auxiliary tasks that combine diversity generation and discriminative representation learning to improve generalization. This method incorporates self-supervised learning for data augmentation alongside supervised contrastive learning to enhance feature diversity. However, this framework 
requires fine-tuning of the trained model for each left-out set of testing participants.
In addition, the authors apply a modified version of the LOOCV benchmark in which a reduced part of the available testing data is evaluated, and this limits direct comparisons with other methods. 
In contrast to the previous works, we propose a new adversarial framework that integrates a new discrimination task that takes into account the inter-person variability for the same activity. In particular, our discrimination task is framed as a binary classification problem that does not scale with the number of participants as in \cite{ShuSungho2022Percom}. Furthermore, and compared to \cite{bai2020adversarial}, we included the activity information in the discrimination task, and we re-structured the original dataset to improve the generalization among all the participants. Finally, our approach integrates a state-of-the-art feature extractor with a reduced parameter count \cite{calatrava2023ieeesensors}.

\section{Activity-based Adversarial Discrimination Framework}
\label{sec:method}

In this section, we explain our adversarial framework for the HAR problem. Our input is composed of signal data points coming from $c$ different sensor cues. For each sensor cue $c$, data points are segmented into subsets using a sliding window approach with a window size of $w$. Therefore, our data input set is defined as:
\begin{align}
X=\{x_i\}_{i=1}^M \ | \  x_i \in R^{w \times c} 
\end{align}
where $M$ is the total number of input data samples in our dataset. We then defined our labeled input dataset set as:
\begin{align}
A = \{x_i,y_i,s_i\}_{i=1}^{M} \  | \ y_i \in Y , s_i \in S
\end{align}
where $Y=\{y_1, \ldots, y_K\}$ represents the set of $K$ activity labels to be recognized, and $S$ is the set of subject participants. 

\begin{figure}[t]
    \centering
        \includegraphics[width=\columnwidth]{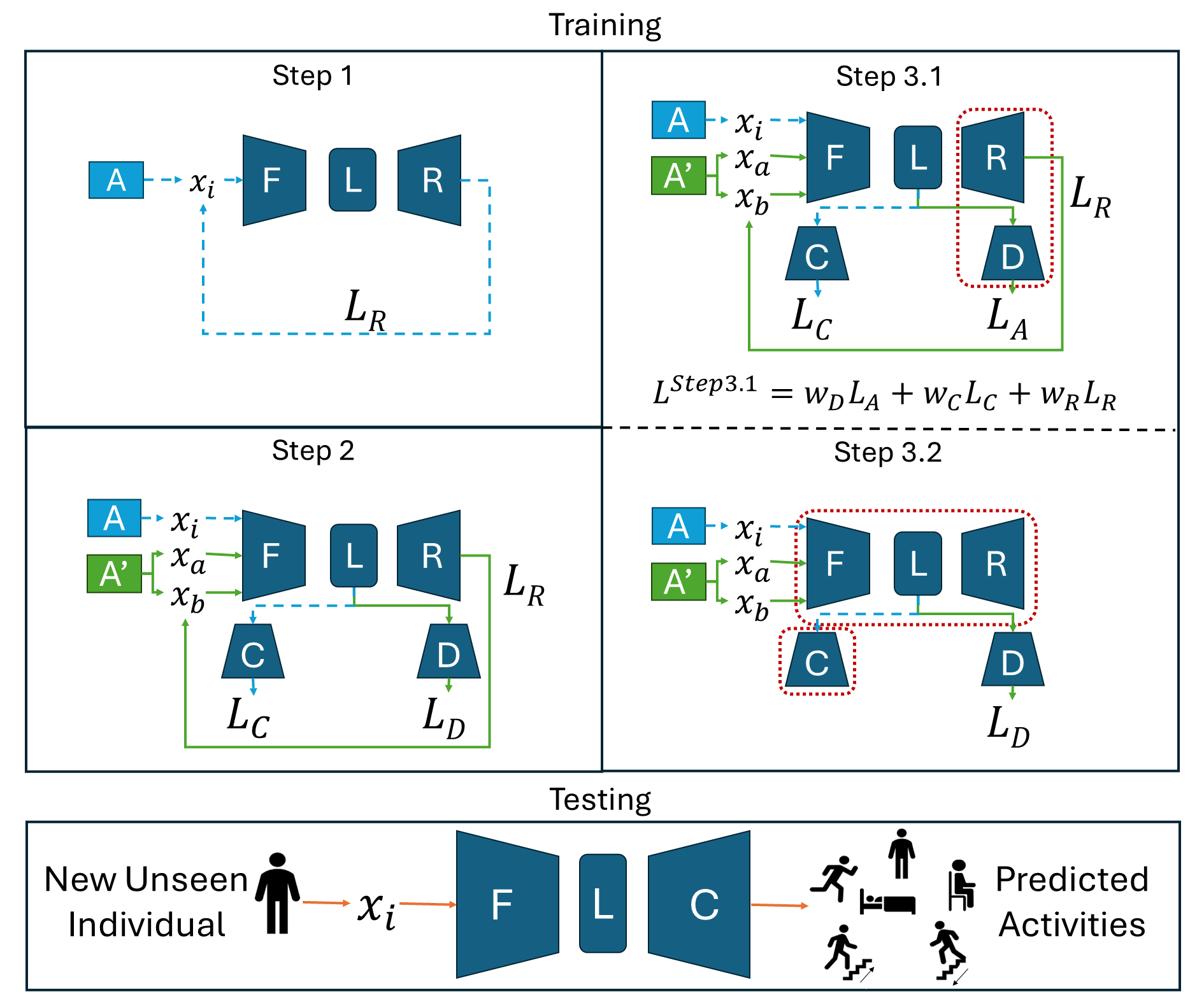}
        \caption{In our adversarial framework blue-dashed lines indicate the path followed by samples in $A$, green-full lines indicate the path for samples in $A'$, and red-dotted lines indicate the frozen parts during the adversarial learning process. The testing phase assesses the activity recognition performance for new, unseen individuals using only the previously trained feature extractor and classifier.
        .}
        \label{fig:architecture}
\end{figure}

In this work, we additionally define a new input dataset $A'$ that will be used in our activity-based discrimination task as:
 \begin{align}
 A' {=} & \{ (x_a,x_b)^y_i, (s_a,s_b)_i, g_i \}_{i=1}^{N}\ | \nonumber \\  
      & x_a, x_b{\in}X,x_a{\ne}x_b, y{\in}Y, s_a,s_b{\in}S, g_i{\in}G{=}\{0,1\}
 \end{align}

\noindent where $(x_a,x_b)^y$ represent two different data samples $x_a$ $x_b \in X$ that share the same activity label $y \in Y$; 
$(s_a, s_b)$ are two subjects from which the corresponding feature vectors $(x_a$, $x_b)$ are selected. Finally, $g_i$ is a binary label with value $1$ if data samples $x_a$ and $x_b$ belong to the same subject, i.e. $s_a = s_b$, or value $0$ if they belong to different subjects, i.e. $s_a \neq s_b$. 

The idea behind $A'$ is to take into account the inter-person variability, that is, different individuals implement the same activity in different ways. 
The size of $A'$ grows with the potential combinations of activities and subjects. To keep a manageable size, we reduce the size of $A'$ using a uniform sampling strategy that keeps an equal number of samples for each label in $G$. In addition, for each label we distribute the samples equally among the number of potential combinations of participants, i.e., for label $g_i=1$ the number of combinations is $|S|$ (same participant), whilst for label $g_i=0$ the combinations are $\binom{|S|}{2}$ (different participants).

\begin{table}[t]
\centering
\caption{Parameters for our reconstructor. After each convolution, there is a batch normalization+LEakyReLU with a negative slope of 0.01. The variable $c$ denotes the number of sensor cues. Legend: OSh (Output Shape), Ker (Kernel), Str (Stride), Pad (Padding), Dil (Dilation), OPa (Output Padding).}
\begin{tabular}{@{}lcccccc@{}}
\toprule
\textbf{Layer} & \textbf{OSh} & \textbf{Ker} & \textbf{Str}& \textbf{Pad}& \textbf{Dil} & \textbf{OPa} \\ 
\midrule
Linear          & [1, 1, 128]      &  -   &   -   &   -   & -  &   -     \\
ConvTranspose2d & [128, 1, 32]     & [1,3] & [1,1] & [0,1] & 16 & [0,1]  \\
Conv2d          & [128, 1, 32]     & [1,3] & [1,1] & [0,1] & -  &   -    \\
ConvTranspose2d & [64, 1, 64]      & [1,3] & [1,1] & [0,1] & 16 & [0,2]  \\
Conv2d          & [64, 1, 64]      & [1,3] & [1,1] & [0,1] & -  &   -    \\
ConvTranspose2d & [$c$, 1, 128]  & [1,5] & [1,2] & [0,2] & 1  & [0,1]  \\
ConvTranspose2d & [$c$, 1, 256]  & [1,5] & [1,2] & [0,2] & 1  & [0,1]  \\
ConvTranspose2d & [$c$, 1, 512]  & [1,5] & [1,2] & [0,2] & 1  & [0,1]  \\
\bottomrule
\end{tabular}
\label{tab:reconstructor_architecture}
\end{table}

\begin{table}[t]
\centering
\caption{Parameters for our binary discriminator. Legend: OSh (Output Shape), Ker (Kernel), Str (Stride), Dr (Dropout).}
\begin{tabular}{@{}lccccc@{}}
\toprule
\textbf{Layer} & \textbf{OSh} & \textbf{Ker} & \textbf{Str} &\textbf{Dr}\\
\midrule
ConvBlock 1                   & [1, 32, 60] & [1,9] & [1,2]  & - \\
ConvBlock 2                   & [1, 64, 28] & [1,5] & [1,2] & - \\
ConvBlock 3                   & [1, 128, 13] & [1,3] & [1,2] & - \\
Dropout                       & [1, 128, 13] & - & -  & 0.3 \\
ConvBlock 4                   & [1, 256, 6] & [1,3] & [1,2]  & - \\
Dropout                       & [1, 256, 6] & - & -  & 0.3\\
Flatten                       & [-1, 1536] & - & -  & -\\
Linear                        & [-1, 256]  & - & -   & -\\
Dropout                       & [-1, 256]  & - & -   & 0.2\\
Linear                        & [-1, 64] & - & - \\
Linear                        & [-1, 1] & -  & - \\
\bottomrule
\end{tabular}
\label{tab:discriminator_architecture}
\end{table}

Our adversarial framework is shown in Fig. \ref{fig:architecture}. Its main components are the feature extractor (F), the reconstructor (R), the activity classifier (C), and the discriminator (D). The feature extractor encodes the input sensor data $\mathcal{X}$ into a lower dimensional latent space $\mathcal{L}$, i.e. $F:\mathcal{X} \rightarrow \mathcal{L}$. 
The objective of $F$ is to identify a latent feature space that, while fooling the discriminator, remains effective for the activity classification task. In this work, we use the network from \cite{calatrava2023ieeesensors} without the classification layer. 

The reconstructor decodes the latent space $\mathcal{L}$ into the original input space $\mathcal{X}$, i.e. $R:\mathcal{L} \rightarrow \mathcal{X}$. This block helps preserve the characteristics from the input signal and stabilizes the adversarial part of the training process.  
The parameters of the reconstructor are shown in Table \ref{tab:reconstructor_architecture}. We use the mean square error function for the reconstruction loss as:

\begin{align}
L_{R} = ||R(F_i)-x_i||_{2}^{2} 
\end{align}

\noindent where $F_i$ stands for  $F(x_i)$. The activity classifier maps the latent space $\mathcal{L}$ into the activities space $\mathcal{Y}$, i.e., $C:\mathcal{L} \rightarrow \mathcal{Y}$. It is composed of 3 fully connected layers of $64 \times 256$, $256 \times 512$, and $512 \times |Y|$ with a ReLU activation function between them. The last layer uses a softmax function to output a probability vector. We use the cross-entropy loss for the activity classifier is defined as:

 \begin{align}
 L_{C} = -\sum\nolimits_{j=1}^{|Y|} z_j \log C(F_i)_j
\end{align}

\noindent where $z$ is a binary vector of size $|Y|$  whose component $z_j=1$ if the original activity label for $x_i$ is $y_j$ or zero otherwise

Finally, our binary discriminator maps the latent space $\mathcal{L}$ into our activity-based binary discriminative classes $G$, i.e. $D:\mathcal{L} \rightarrow \mathcal{G}$. The discriminator is composed of 4 convolutional blocks made of convolution + batch normalization + ReLU followed by fully connected layers. Specific parameters for our binary discriminator are shown in Table \ref{tab:discriminator_architecture}. We use a binary cross-entropy loss for the discriminator as:


\begin{align}
    \resizebox{.89\hsize}{!}{$
    L_{\scriptscriptstyle D} = -g_{\scriptscriptstyle i} \cdot \log D\left( (F_{\scriptscriptstyle a}, F_{\scriptscriptstyle b})_{\scriptscriptstyle i} \right) + (1 - g_{\scriptscriptstyle i}) \cdot \log 1 - D\left( (F_{\scriptscriptstyle a}, F_{\scriptscriptstyle b})_{\scriptscriptstyle i} \right)
    $}
    \end{align}
    
\noindent where $F_{a}$ and $F_{b}$ stand for $F(x_a)$ and $F(x_b)$ respectively, $x_a$ and $x_b$ represent the two data samples comprising the $i_{th}$ pair, and $g_i$ corresponds to the label of the $i_{th}$ pair.

\begin{algorithm}[tb]
\caption{Training algorithm}
\begin{algorithmic}[1] 
    \State ------------------------------Step 1--------------------------------
    \For{$epoch = 1$ to $epochs$}
        \For{each $batch$ in $X_A$ from $A$}
            \State $\theta_{F} \leftarrow \theta_{F} - \grad\theta_{F} \cdot \alpha_{F}\ L_{F}^{step1}(X_A,\theta_{F})$ 
            \State $\theta_{R} \leftarrow \theta_{R} - \grad\theta_{R} \cdot \alpha_{R}\ L_{R}^{step1}(X_A,\theta_{R})$
        \EndFor
    \EndFor
    \State ------------------------------Step 2--------------------------------
    \For{$epoch = 1$ to $epochs$}
        \For{each $batch$ in $X_A,Y_A$ and $X_{A'}, G_{A'}$ from $A$ and $A'$}
            \State $\theta_{F} \leftarrow \theta_{F} - \grad\theta_{F} \cdot \alpha_{F}\ L_{F}^{step2}(X_A,X_{A'},\theta_{F})$
            \State $\theta_{R} \leftarrow \theta_{R} - \grad\theta_{R} \cdot \alpha_{R}\ L_{R}^{step2}(X_{A'},\theta_{R})$
            \State $\theta_{C} \leftarrow \theta_{C} - \grad\theta_{C} \cdot \alpha_{C}\ L_{C}^{step2}(X_A,\theta_{C})$
            \State $\theta_{D} \leftarrow \theta_{D} - \grad\theta_{D} \cdot \alpha_{D}\ L_{D}^{step2}(X_{A'},\theta_{D})$
        \EndFor
    \EndFor
    \State ------------------------------Step 3--------------------------------
    \State We freeze R
    \For{$epoch = 1$ to $epochs$}
        \For{each $batch$ in $X_A,Y_A$ and $X_{A'}, G_{A'}$ from $A$ and $A'$}
            \State We freeze D
            \State $\theta_{F} \leftarrow \theta_{F} - \grad\theta_{F} \cdot \alpha_{F}\ L_{F}^{step3.1}(X_A,X_{A'},\theta_{F})$
            \State $\theta_{C} \leftarrow \theta_{C} - \grad\theta_{C}\cdot \alpha_{C}\ L_{C}^{step3.1}(X_A,\theta_{C})$
            \State We unfreeze D and we freeze F and C
            \State $\theta_{D} \leftarrow \theta_{D} - \grad\theta_{D} \cdot \alpha_{D}\ L_{D}^{step3.2}(X_{A'},\theta_{D})$
        \EndFor
    \EndFor
\end{algorithmic}
\label{alg:training}
\end{algorithm}
Drawing inspiration from \cite{motiian2017few} and \cite{ShuSungho2022Percom}, our training process is divided into three steps. The initial two steps focus on pre-training the weights, while the third step is dedicated to adversarial learning. More specifically, in step 1 we only train $F$ and $R$ by applying the following loss function:
\begin{align}
    L_F^{step1} = L_R^{step1} = L_R
\end{align}
In step 2 we train the framework in a multi-task way by training all the models of our framework at the same time. In this step, $F$ updates its parameters based on the reconstruction and classification losses, $R$ updates its parameters based on the reconstruction loss, $C$ updates its parameters based on the classification loss, and $D$ updates its parameters based on the discriminator loss as follows:
\noindent
\begin{minipage}{.5\linewidth}
\begin{align}
    L_F^{step2} &= L_R + L_C\\
    L_R^{step2} &= L_R
\end{align}
\end{minipage}%
\begin{minipage}{.5\linewidth}
\begin{align}
    L_C^{step2} &= L_C\\
    L_D^{step2} &= L_D
\end{align}
\end{minipage}\\

Step 3 is the adversarial process that is subdivided into two further sub-steps. In the first sub-step, $F$ is trained based on the reconstruction, discriminator, and classification losses, while $R$ and $D$ are frozen. In the second sub-step, the parameters from $D$ are updated based on the discriminator loss, while $R$, $F$, and $C$ are frozen:
\begin{align}
    L_F^{step3.1} &{=} L_C^{step3.1} = w_D * L_A + w_R * L_R + w_C * L_C \\
    L_D^{step3.2} &{=} L_D
    \label{eq:loss3.1}
\end{align}

\noindent where $w_D$, $w_R$, and $w_C$ represent weights, and $L_A$ is the non-saturating GAN loss function \cite{goodfellow2016nips} for the adversarial part:

\begin{align}
    L_{\scriptscriptstyle A}{=}-\log D(F_{\scriptscriptstyle a},F_{\scriptscriptstyle b})), \forall \{(x_{\scriptscriptstyle a},x_{\scriptscriptstyle b})_{\scriptscriptstyle i}, g_{\scriptscriptstyle i}\}{\in} A' {\mid} g_{\scriptscriptstyle i}{=}0
    \label{eq:Loss_A}
\end{align}

The algorithm for the full training process is shown in Algorithm \ref{alg:training}. Please note that, as shown in Fig. \ref{fig:architecture}, in the testing phase only the feature extractor and the classifier are utilized.
The other components of the framework, such as the reconstructor and discriminator,
are only employed during training and validation. During training, these components push the feature extractor to reach a deep feature space that deceives
the discriminator. Validation is used to monitor for overfitting. In testing, our
focus is solely on the feature extractor and the activity classifier.

\section{Experiments}

\begin{table}[t]
\centering
\caption{Hyperparameters}
\begin{tabular}{llccc}
\toprule
& & Learning Rate    & Epochs & Optimizer \\
\midrule
Step 1 & $F$ & le-4  & 15     & Adam \\
       & $R$ & le-4  & 15     & Adam \\
       \midrule
Step 2 & $F$ & le-4  & 5      & Adam \\
 & $R$ & le-4  & 5      & Adam \\
 & $C$ & le-4  & 5      & Adam \\
 & $D$ & le-3  & 5      & Adam \\
 \midrule
Step 3 & $F$ & le-4  & 150    & Adam \\
 & $C$ & le-4  & 150    & Adam \\
 & $D$ & le-4  & 150    & Adam \\
\bottomrule
\end{tabular}
\label{tab:hyperparameters}
\end{table}

In this section, we first describe the datasets used in our experiments and the parameters for our model. We then present a comparison of classification results with previous approaches. We also compare our activity-based discrimination task with tasks proposed in previous works.

\subsection{Datasets}
\label{sec:datasets}

We apply our adversarial method to three activity datasets: PAMAP2 \cite{Reiss2012}, MHEALTH \cite{banos2014mhealthdroid}, and REALDISP  \cite{realdisp}. 
The PAMPA2 dataset \cite{Reiss2012} contains data from 18 different physical activities carried out by 9 subjects while wearing 3 inertial measurement units (IMUs) and a heart rate sensor. The IMUs are placed over the wrist on the dominant arm, on the chest, and on the dominant leg. Each subject followed a protocol composed of 12 activities: lying, sitting, standing, walking, running, cycling, nordic walking, ascending stairs, descending stairs, vacuum cleaning, ironing, and rope jumping, which are labelled from 0 to 11 in the listed order.
A total of 54 signal cues are available with the corresponding timestamps and labels. In our experiments, we classify the 12 activities from the protocol. We use only the data from the inertial sensor cues (3D acceleration with the scales of +-16g and gyroscope), i.e. a total of 18 signal cues. As in previous works, the data from the ninth participant is not used due to a lack of sufficient samples. We segment the data with a sliding window approach with a window size of 512 samples and an overlap of 256 samples. 

The MHEALTH dataset \cite{banos2014mhealthdroid} contains data of 12 different activities from 10 participants that wore 3 different sensors. Sensors are located on the chest, the right wrist, and the left ankle. Each sensor is composed of an IMU providing 3D acceleration, a gyroscope, and a magnetometer, except the sensor from the chest, which only provides 3D acceleration and ECG. The 12 activities are standing, sitting, lying, walking, climbing stairs, waist bends, frontal elation of arms, knees bending, cycling, jogging, running, jump front and back; which are labelled from 0 to 11 in the listed order. We segment the data using a sliding window approach with a window size of 512 samples and an overlapping of 256 samples. 

The REALDISP dataset \cite{realdisp} contains data of 33 different activities from 17 participants wearing 9 different sensors, that are located on the left calf, left thigh, right calf, right thigh, back, left lower arm, left upper arm, right lower arm, and right upper arm. Each sensor provides 3D acceleration, 3D gyroscope, 3D magnetic field orientation, and 4D quaternions.  We utilize data from both the 3D accelerometer and the gyroscope, resulting in a total of 54 signals. The 33 activities are walking, jogging, running, jumping up, jumping front and back, jumping sideways, jumping leg/arms open/closed, jumping rope, trunk twist (arms outstretched), trunk twist (elbows bent), waist bending forward, waist rotation, waist bends (reach foot with opposite hand), reach heels backwards, lateral bend (left and right), lateral bend with arm up (left and right), repetitive forward stretching, upper trunk and lower body opposite twist, lateral elevation of arms, frontal elevation of arms, frontal hand claps, frontal crossing of arms, shoulders high-amplitude rotation, shoulders low-amplitude rotation, arms inner rotation, knees (alternating) to the breast, heels (alternating) to the backside, knees bending (crouching), knees (alternating) bending forward, rotation on the knees, rowing, elliptical bike, and cycling; which are labelled from 0 to 32 in the listed order. We segment the data using a sliding window approach with a window size of 256 samples and an overlapping of 128 samples. 
Following the suggestions in \cite{hammerla2015let,calatrava2023ieeesensors}, for each dataset, we apply a LOOCV approach where we designated one participant for testing, two for validation, and the rest for training. We ensured that the validation set included at least two participants to monitor whether the discriminator was fooled. The testing phase focused on evaluating the activity classification task using the feature extractor and classifier trained in the adversarial framework, which is why pairs of signals are not required during this testing phase. In addition, we normalize the datasets using a min-max method based on the parameters obtained from the training sets. To address the small variability between runs we repeated each test twice and calculated the average.

\begin{table*}[t]
\centering
\caption{Comparison with previous approaches. Best result in bold.}
\begin{tabular}{lllcccccc}
\toprule
Dataset & & Model & &  Accuracy & & $F1-Score_{M}$ & & $F1-Score_{W}$ \\
\midrule
 
 &&MC-CNN \cite{yang2015deep}  & & $0.7939 \pm 0.0770$ & & $0.7511 \pm 0.0940$ & & $0.7801 \pm 0.0906$ \\
 && DCLSTM \cite{ordonez2016deep} & & $0.7853 \pm 0.1012$ & & $0.7266 \pm 0.1113$ & & $0.7708 \pm 0.1159$ \\
 PAMAP2 \cite{Reiss2012}&& METIER \cite{chen2020metier} & & $0.7924 \pm 0.0597$ & & $0.7540 \pm 0.0574$ & & $0.7888 \pm 0.0611$ \\
 && UIDFE \cite{ShuSungho2022Percom} & & $0.8014 \pm 0.1353$ & & $0.7648 \pm 0.1447$ & & $0.7883 \pm 0.1536$ \\
 && DDLearn \cite{qin2023generalizable} &  & $0.7188 \pm 0.1561$ & & $0.6442 \pm 0.1691$ & & $0.7097 \pm 0.1776$\\
  && \textbf{Ours} & & $\mathbf{0.8703 \pm 0.1219}$ & & $\mathbf{0.8643 \pm 0.1431}$ & & $\mathbf{0.8609 \pm 0.1437}$ \\
\midrule
 && MC-CNN \cite{yang2015deep} & & $0.8736 \pm 0.0739$ & & $0.7888 \pm 0.0823$ & & $0.8376 \pm 0.0931$ \\
 && DCLSTM \cite{ordonez2016deep} & & $0.8482 \pm 0.1051$ & & $0.8157 \pm 0.1169$ & & $0.8196 \pm 0.1173$ \\
 MHEALTH \cite{banos2014mhealthdroid}&& METIER \cite{chen2020metier} & & $0.8583 \pm 0.0746$ & & $0.8252 \pm 0.0932$ & & $0.8317 \pm 0.0924$ \\
 && UIDFE \cite{ShuSungho2022Percom}& & $0.8982 \pm 0.0667$ & & $0.8275 \pm 0.0881$ & & $0.8685 \pm 0.0867$ \\
 && DDLearn \cite{qin2023generalizable} & & $0.8529 \pm 0.0684$ & & $0.7825 \pm 0.0764$ & & $0.8263 \pm 0.0773$ \\
 && \textbf{Ours} & & $\mathbf{0.9225 \pm 0.0606}$ & & $\mathbf{0.9065 \pm 0.0780}$ & & $\mathbf{0.9057 \pm 0.0786}$ \\
\midrule
 && MC-CNN \cite{yang2015deep} & & $0.8378 \pm 0.0601$ & & $0.6651 \pm 0.0998$ & & $0.7916 \pm 0.0757$ \\
 && DCLSTM \cite{ordonez2016deep} & & $0.9420 \pm 0.0497$ & & $0.9094 \pm 0.0627$ & & $0.9308 \pm 0.0600$ \\
 REALDISP \cite{realdisp}&& METIER \cite{chen2020metier} & & $0.9177 \pm 0.0567$ & & $0.8099 \pm 0.1080$ & & $0.9038 \pm 0.0690$ \\
 && UIDFE \cite{ShuSungho2022Percom}& & $0.9450 \pm 0.0670$ & & $0.9283 \pm 0.0700$ & & $0.9366 \pm 0.0816$ \\
 && DDLearn \cite{qin2023generalizable} & & $0.8476 \pm 0.0902$ & & $0.7802 \pm 0.1323$ & & $0.8296 \pm 0.0995$ \\
 && \textbf{Ours} & & $\mathbf{0.9710 \pm 0.0423}$ & & $\mathbf{0.9651 \pm 0.0368}$ & & $\mathbf{0.9658 \pm 0.0513}$ \\
\bottomrule
\end{tabular}
\label{tab:ComparisonSOTA}
\end{table*}

\begin{table*}[t]  
\centering
\caption{Comparison with previous discrimination tasks. Best result in bold. }
\begin{tabular}{lllcccccc}
\toprule
 Dataset & & Discriminator & & Accuracy & & $F1-Score_{M}$ & & $F1-Score_{W}$ \\
\midrule
\multirow{3}{*}{PAMAP2 \cite{Reiss2012}}
 && $D_{i}$ & & $0.8274 \pm 0.1045$ & & $0.7869 \pm 0.1102$ & & $0.8157 \pm 0.1194$ \\
 && $D_{b}$ & & $0.8130 \pm 0.1453$ & & $0.7636 \pm 0.1579$ & & $0.8046 \pm 0.1585$ \\
 && \textbf{Ours} & & $\mathbf{0.8703 \pm 0.1219}$ & & $\mathbf{0.8643 \pm 0.1431}$ & & $\mathbf{0.8609 \pm 0.1437}$ \\
\midrule
\multirow{3}{*}{MHEALTH \cite{banos2014mhealthdroid}}
 && $D_{i}$ & & $0.9173 \pm 0.0698$ & & $0.8892 \pm 0.0938$ & & $0.8955 \pm 0.0901$ \\
 && $D_{b}$ & & $0.9225 \pm 0.0707$ & & $0.9010 \pm 0.0867$ & & $0.9037 \pm 0.0913$ \\
&& \textbf{Ours} & & $\mathbf{0.9225 \pm 0.0622}$ & & $\mathbf{0.9065 \pm 0.0801}$ & & $\mathbf{0.9057 \pm 0.0807}$ \\
\midrule
\multirow{3}{*}{REALDISP \cite{realdisp}}
 && $D_{i}$ & & $0.9447 \pm 0.0772$ & & $0.9394 \pm 0.0566$ & & $0.9386 \pm 0.0865$ \\
 && $D_{b}$ & & $0.9511 \pm 0.0619$ & & $0.9445 \pm 0.0538$ & & $0.9424 \pm 0.0755$ \\
&& \textbf{Ours} & & $\mathbf{0.9710 \pm 0.0423}$ & & $\mathbf{0.9651 \pm 0.0368}$ & & $\mathbf{0.9658 \pm 0.0513}$ \\
\bottomrule
\end{tabular}
\label{tab:discriminator_comparison}
\end{table*}

\subsection{Setup}
\label{sec:setup}

We compare our approach with previous works on HAR using inertial sensors, in particular, the Multichannel Deep Neural Network (MC-CNN) from \cite{yang2015deep}, the deep convolutional LSTM (DCLSTM) method from \cite{ordonez2016deep}, the deep multi-task learning based activity and user recognition (METIER) approach from \cite{chen2020metier}, the user-invariant deep feature extractor (UIDFE) from \cite{ShuSungho2022Percom}, and the Diverse and Discriminative representation Learning (DDLearn) from \cite{qin2023generalizable}. 

It is important to note that the original UIDFE method incorporates unlabeled test data into the training process and thus, assumes that test data is available in advance. On the contrary, we assume no availability of any test data in advance and, therefore, we have modified the UIDFE method from \cite{ShuSungho2022Percom} to avoid the use of test data in the training process while the rest of the method remains the same. In addition, testing datasets in DDLearn \cite{qin2023generalizable} contains 
only 20\% of the left-out testing participants. In our work, however, we use 100\% of the testing participants. This is done to be consistent with the LOOCV benchmarking and to facilitate comparisons with previous works.

We have implemented the previous approaches \cite{yang2015deep,ordonez2016deep,chen2020metier,ShuSungho2022Percom} following the original papers\footnote{Our code will be publicly available after acceptance of this paper.}. For the approach in 
\cite{qin2023generalizable} we used directly their available code\footnote{https://github.com/microsoft/robustlearn/tree/main/ddlearn}.
The methods were implemented in Python3.10.12 using Pytorch2.2.0+cu121 as a deep learning framework. All the training and testing processes were carried out on Ubuntu 18.04.6 LTS with two NVIDIA Quadro RTX 6000 GPUs. The hyperparameters for our framework are presented in Table \ref{tab:hyperparameters}.
The values for the weights $w_R=0.7$, $w_C=0.2$, and $w_D=0.1$ in Eq. \ref{eq:loss3.1} were found
using a grid search approach. Under the same set of hyper-parameters, we run all the experiments with the same conditions of preprocessing and segmentation steps (cf. Sec. \ref{sec:datasets}). 
In addition, the distribution $A'$ has been resized for each dataset according to their total number of samples. For the PAMAP2, $A'$ contains 50,000 samples, with 25,000 samples for each binary class in $G=\{0, 1\}$. For MHEALTH, $A'$ consists of 10,000 samples, with 5,000 samples for each binary class in $G=\{0, 1\}$. Finally, for REALDISP $A'$ contains a total of 50000 samples, with 25,000 samples for each binary class in $G=\{0, 1\}$. The batch sizes for PAMAP2 were set to 64 for $A$ and 350 for $A'$. For MHEALTH the batch sizes were set to 32 for $A$ and 375 for $A'$. And for REALDISP they are 30 for $A$ and 395 for $A'$.

\section{Comparison with previous approaches}
 \begin{figure*}[t]
     \centering
     \includegraphics[width=0.30\textwidth]{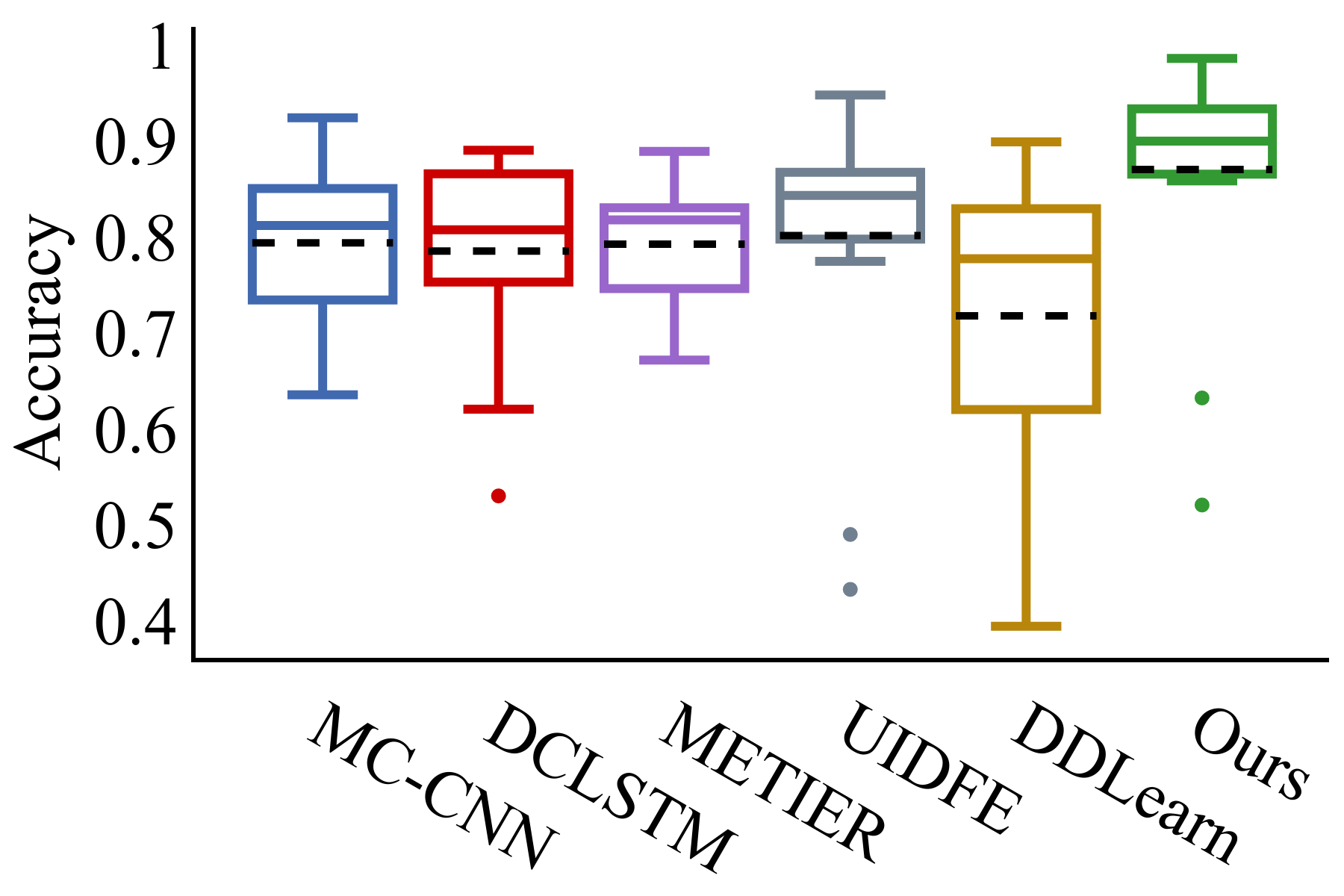}
     \hfill
     \includegraphics[width=0.30\textwidth]{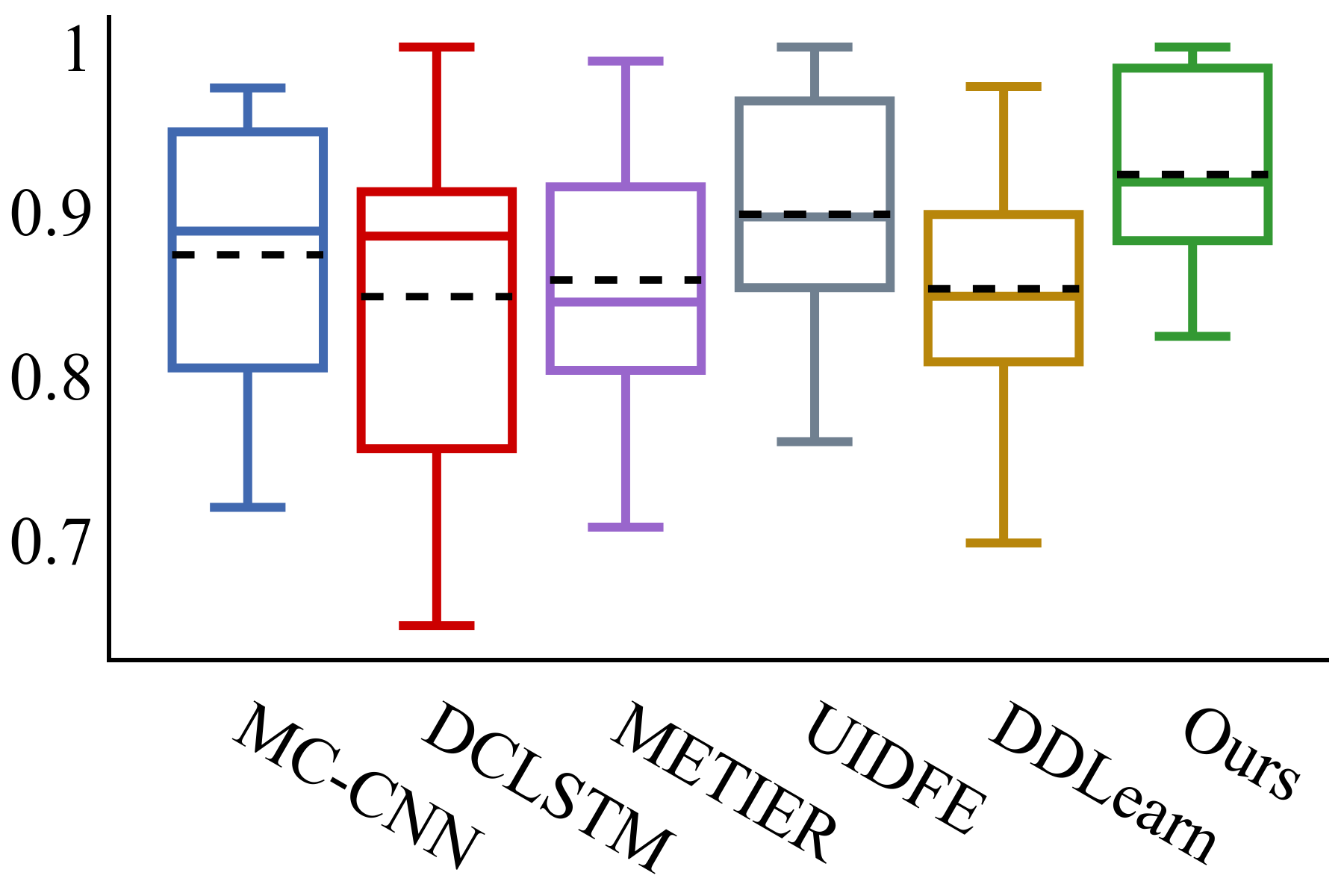}
     \hfill
     \includegraphics[width=0.30\textwidth]{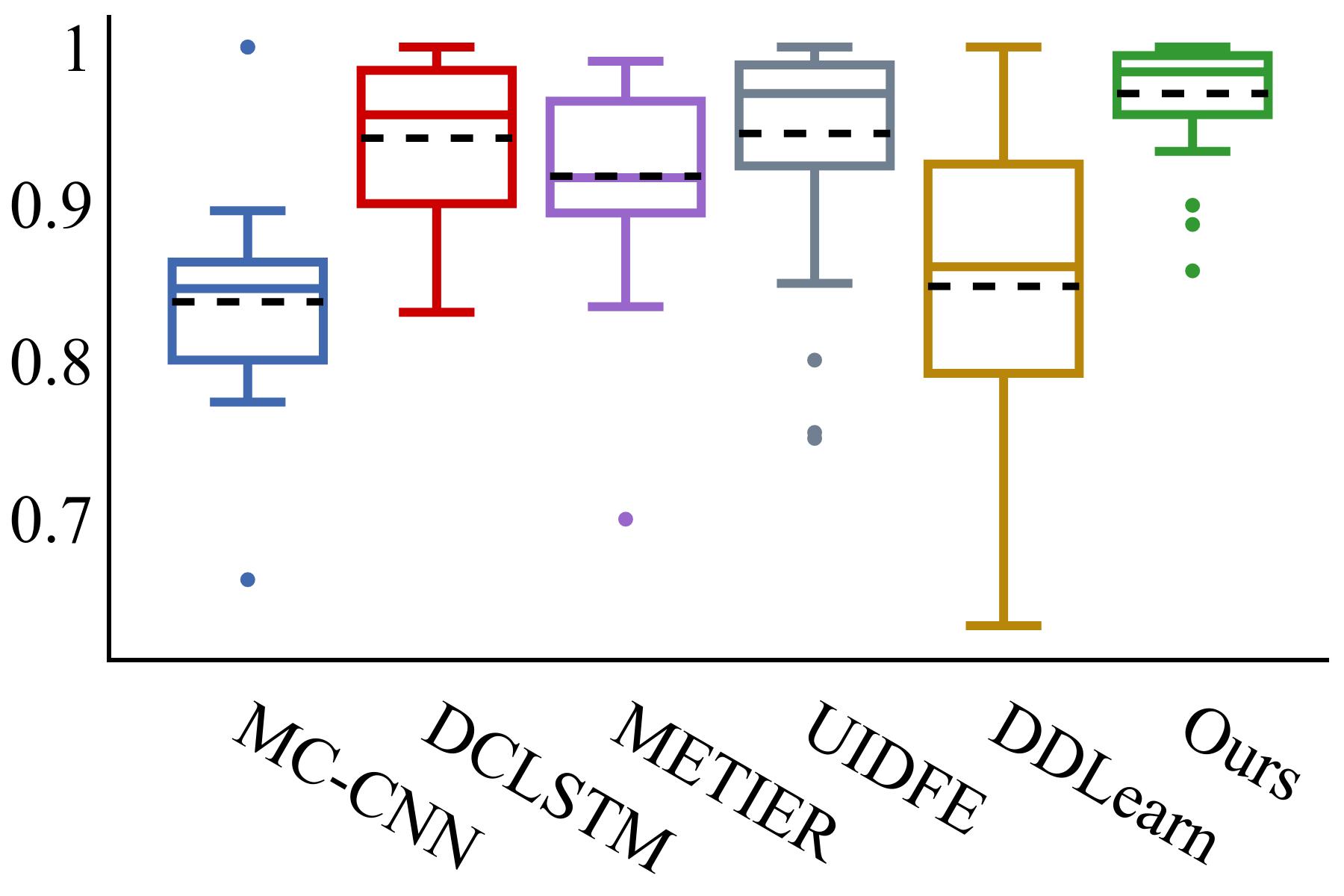}
     \\
     \vspace{7mm}
     \includegraphics[width=0.30\textwidth]{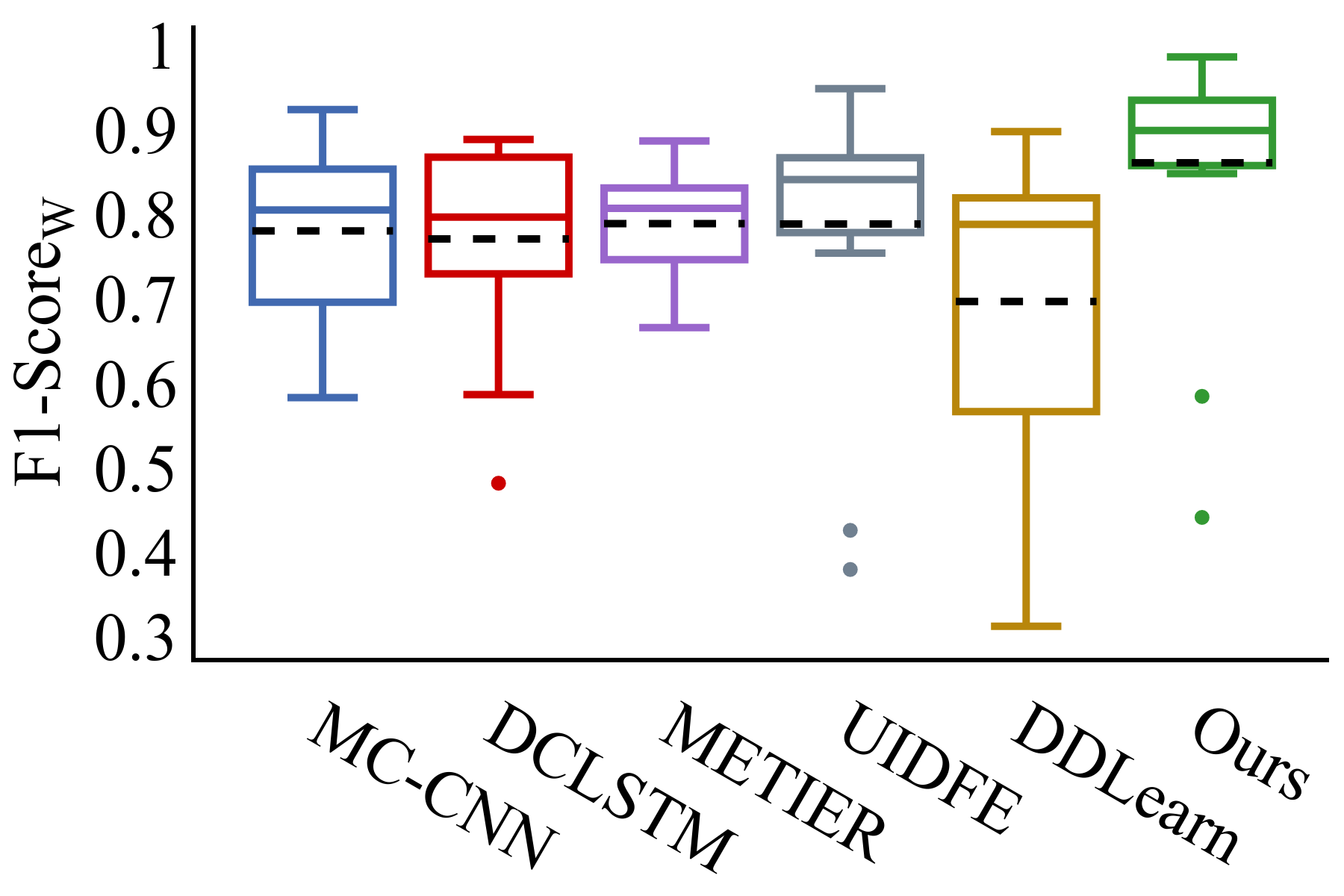}
     \hfill
     \includegraphics[width=0.30\textwidth]{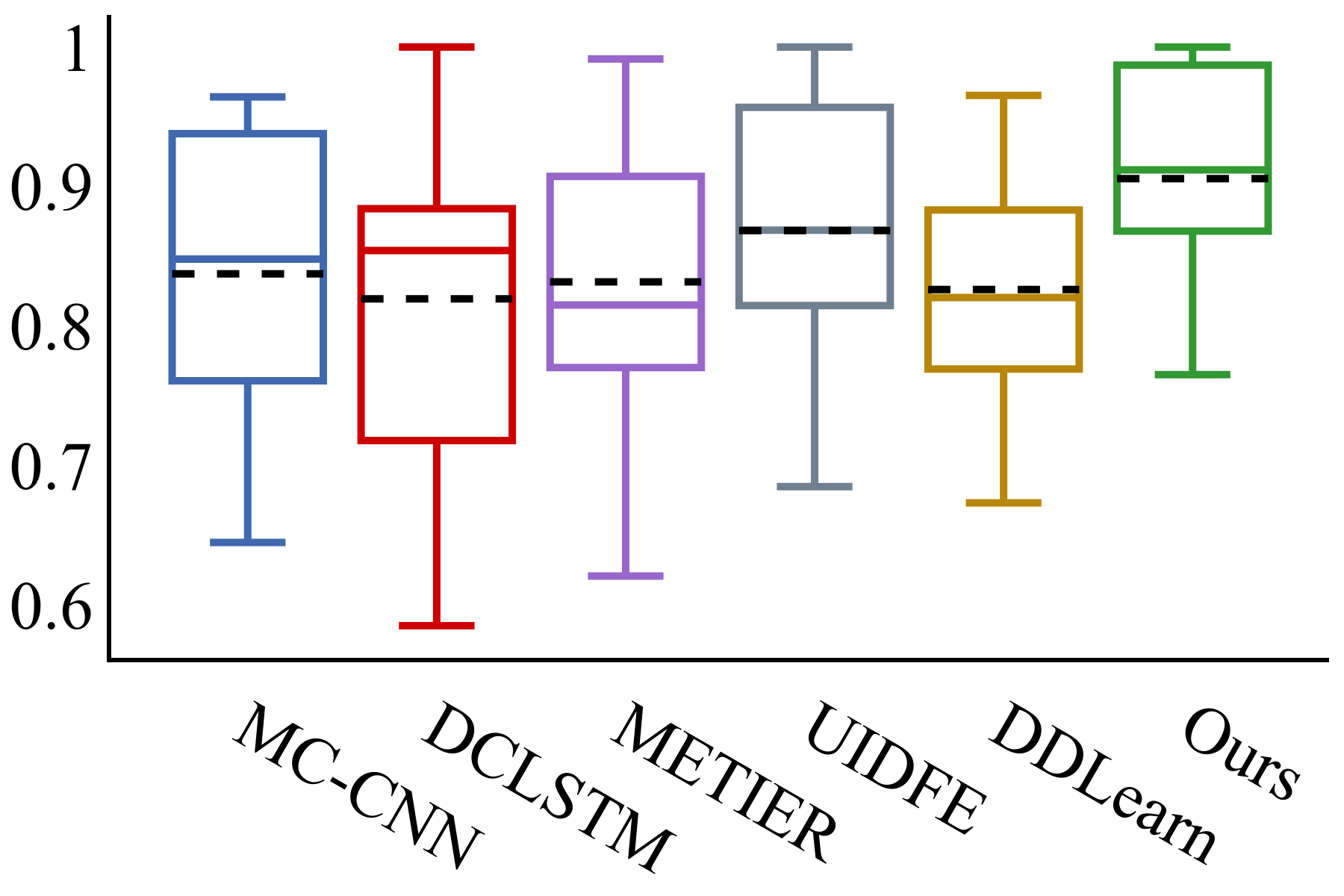}
     \hfill
     \includegraphics[width=0.30\textwidth]{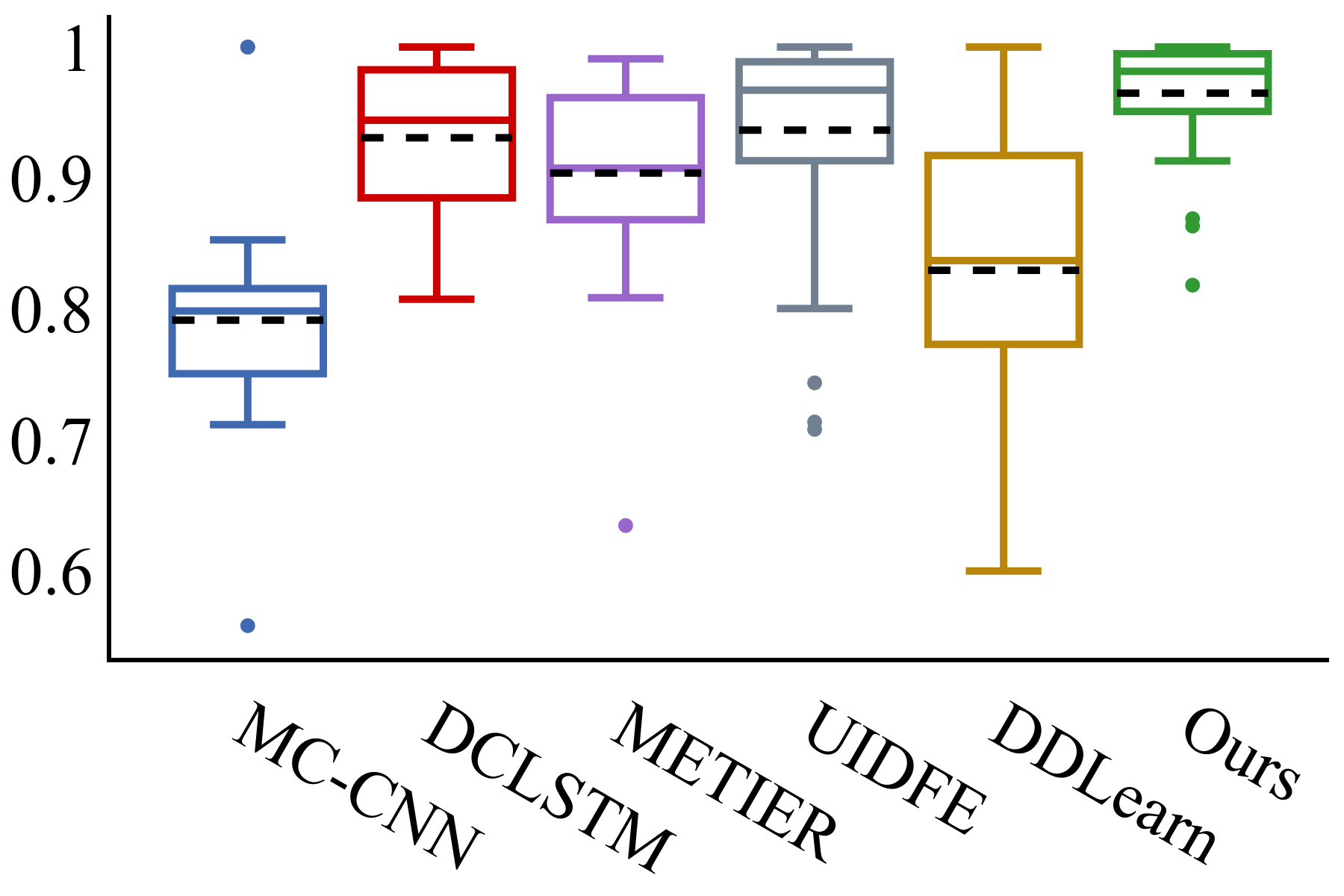}
     \caption{The upper row shows Accuracy while the lower row depicts $F1-Score_{W}$ for datasets PAMAP2 (left), MHEALTH (center), and REALDISP (right). The black dashed lines in the boxes represent the average value.}
     \label{fig:classification_results_state_of_art}
 \end{figure*}

\begin{figure*}[t]
    \centering
    \includegraphics[width=0.30\textwidth]{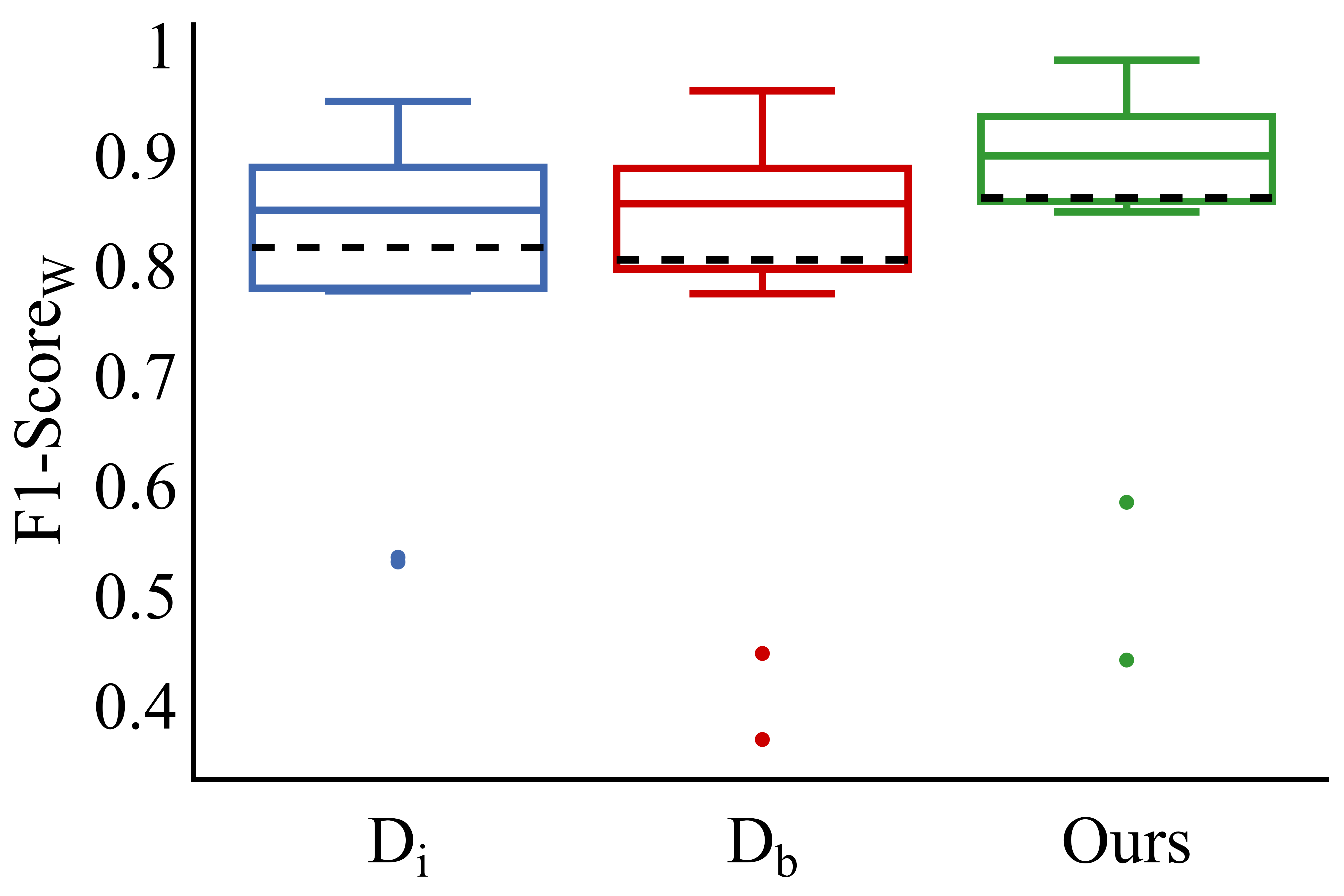}
    \hfill 
    \includegraphics[width=0.30\textwidth]{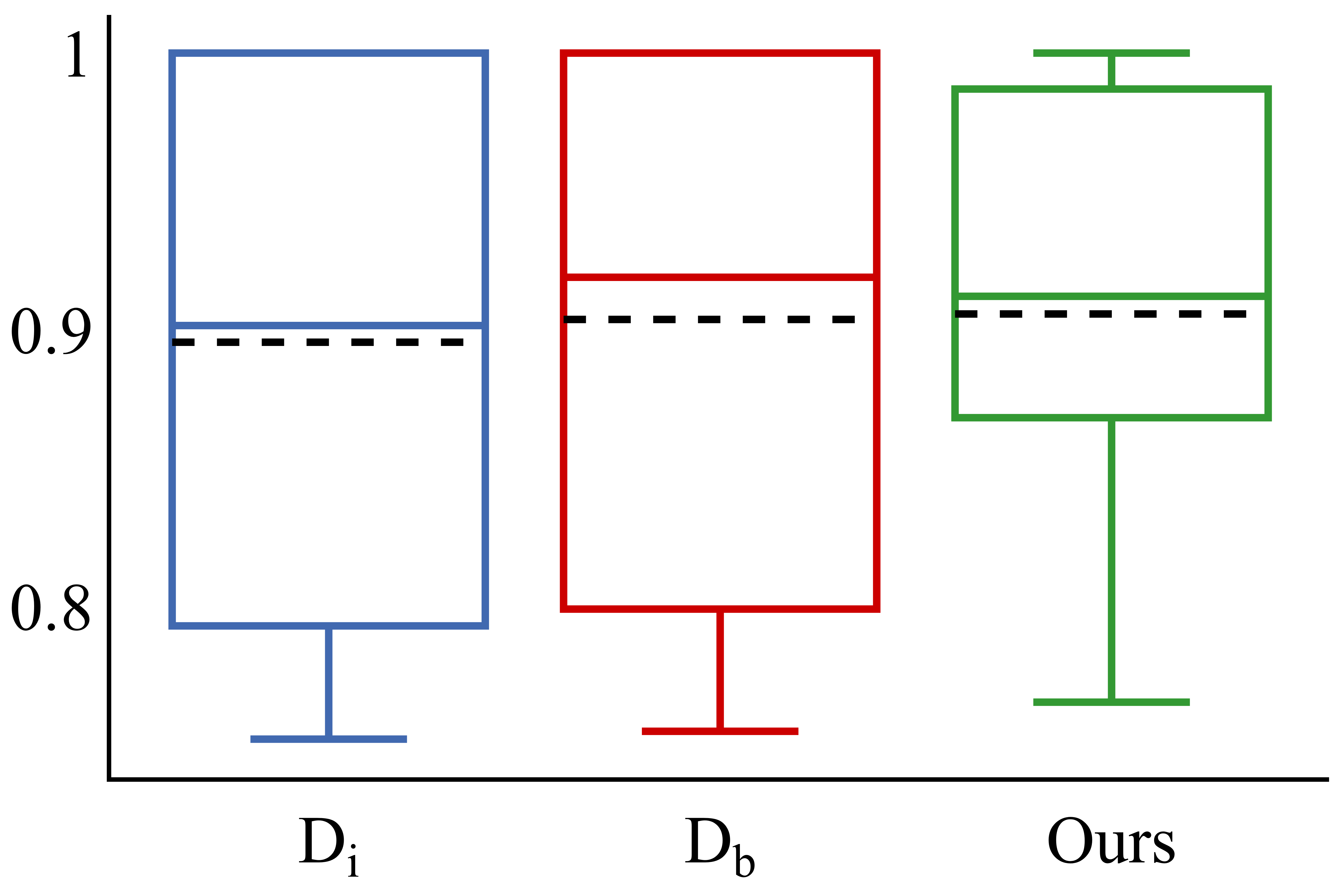}
    \hfill
    \includegraphics[width=0.30\textwidth]{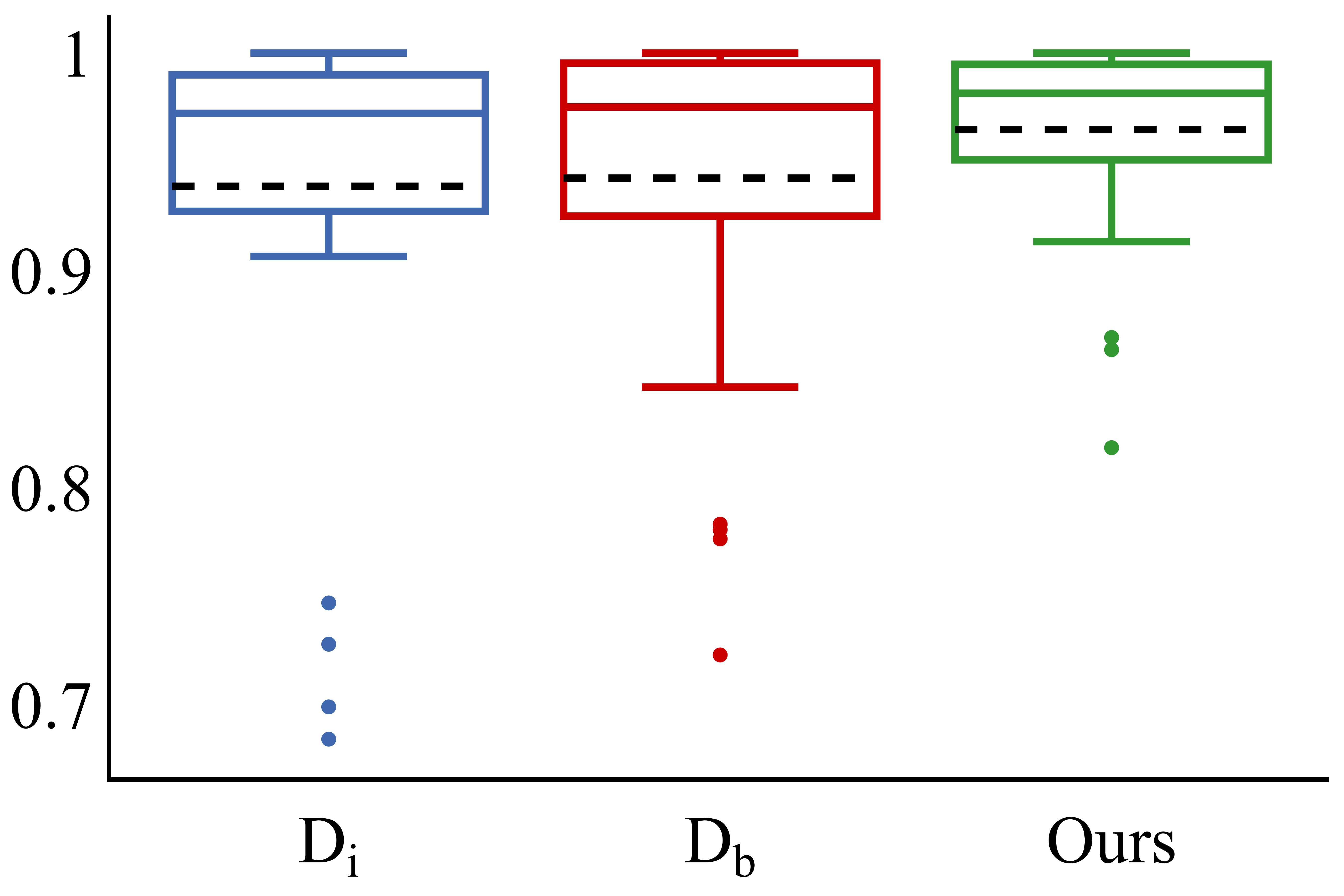}
    \caption{$F-Score_W$ for the different discrimination tasks in PAMAP2 (left), MHEALTH (center), and REALDISP (right). The black dashed lines in the boxes represent the average value.}
    \label{fig:disc_box_plots}
\end{figure*}

We compare our method with the works from \cite{yang2015deep,ordonez2016deep,chen2020metier,ShuSungho2022Percom,qin2023generalizable} (cf. Sec. \ref{sec:setup}) for the metrics Accuracy, F1-Score Macro ($F1-Score_{M}$), and F1-Score weighted ($F1-Score_{W}$) using the Scikit-Learn package \cite{scikit-learn}.
Results are shown in Table \ref{tab:ComparisonSOTA} where  our method outperforms all previous models under the same conditions of pre-processing and segmentation. Moreover, our approach reduces most of the differences between the Accuracy and the two $F1-Score$ metrics (except the difference with $F1-Score_W$ using the METIER model in the PAMAP2 dataset).
This trend suggests that our method remains more stable across diverse dataset compositions, and it demonstrates its ability to effectively handle classes with fewer samples without being overly influenced by the presence of dominant majority classes. 

We also show results using box plots in Figure \ref{fig:classification_results_state_of_art}, where, for the PAMAP2 dataset, our method holds the smaller Interquartile Range (IQR) which translates into a smaller variability. 
In addition, both Q0 and Q4 are the highest for our method indicating a better robustness. Finally, observing the plots for PAMAP2 in Figure \ref{fig:classification_results_state_of_art}, we find one participant in the PAMAP2 dataset that generates significantly lower classification results with all the approaches. These points correspond to the experiments that left out Participant 8. In the PAMAP2 dataset, the sensors on the wrist and the ankle are positioned on the dominant side of the participant, which is usually the right hand. Therefore, it may be the case that this outlier is due to Participant 8 being the only left-handed in the PAMAP2 dataset. To get more insights, we repeated the experiments removing Participant 8 from the PAMAP2 dataset. Still, our method outperform the previous works.

For the MHEALTH dataset, our method not only improves the average classification performance in all the metrics but also reduces the standard deviation. 
In Figure \ref{fig:classification_results_state_of_art} we show that 
the Q0 and the minimum value are the highest. 
Finally, in the REALDISP dataset, we improve the classification performance while reducing the standard deviation across all the metrics, and obtain the best IQR. 
In addition, our minimum value is always higher than any other Q0, showing a very reduced variability.\\
All those results suggest that our method improves robustness and generally provides less variability in the classification.

\subsection{Comparing Different Adversarial Discrimination Tasks}
One of the main contributions of this paper is the activity-based adversarial discrimination task that integrates
the concept of inter-person variability (Sec. \ref{sec:method}). In order to test its performance, we compare it to the discrimination tasks previously presented in \cite{ShuSungho2022Percom} and  \cite{bai2020adversarial}. To this end, we use our proposed framework, in which we keep the same $F$, $R$, and $C$, and we change the discriminator $D$ 
by implementing different discrimination tasks. We first implement a discriminator, $D_i$, that tries to identify each particular participant as proposed in \cite{ShuSungho2022Percom}. This discriminator outputs one class for each participant. The second implemented discriminator, $D_b$, takes as input two feature vectors and decides whether they belong to the same participant or not, as proposed in \cite{bai2020adversarial}. The third discriminator is our activity-based task. 

Comparison results are shown in Table \ref{tab:discriminator_comparison} 
where our discrimination task achieves the best classification results across all datasets. Moreover, it provides the smallest deviation for MHEALTH and REALDISP, and second lowest for PAMAP2.
 
Figure \ref{fig:disc_box_plots} provides additional details using box plots. As shown across all three datasets, our approach is characterized by a smaller IQR, indicating reduced variability between the first and third quartiles ($Q_1$ and $Q_3$). Additionally, the lower quartile ($Q_0$) is higher in our approach for all datasets. 
These results suggest that our proposed activity-based discrimination task generally offers less variability and greater robustness and, 
consequently, we believe it is the best choice when facing the problem of HAR using adversarial learning.

\section{Conclusions}
We presented a new adversarial deep learning model for the problem of human activity recognition on people wearing inertial sensors. Based on the concept of inter-person variability, we designed a new discrimination task that takes into account information about the activity label, with the purpose of finding a common deep feature space for the same activity among different subjects to reduce their distribution distances. 
Our experimental results outperformed previous models in three activity datasets when using a LOOCV benchmark. An additional comparison with previously proposed discrimination tasks showed that our activity-based discrimination task improves the classification results when integrated in the adversarial model.
Future work include testing our model on larger datasets, and implementing cross-dataset validation to better understand its robustness and generalization capabilities.

\bibliographystyle{plain}
\bibliography{IEEEabrv,bibliography}

\begin{thebibliography}{10}

\bibitem{bai2020adversarial}
Lei Bai, Lina Yao, Xianzhi Wang, Salil~S Kanhere, Bin Guo, and Zhiwen Yu.
\newblock Adversarial multi-view networks for activity recognition.
\newblock {\em Proceedings of the ACM on Interactive, Mobile, Wearable and
  Ubiquitous Technologies}, 4(2):1--22, 2020.

\bibitem{realdisp}
Banos, Toth Oresti, Amft Mate, and Oliver.
\newblock {REALDISP Activity Recognition Dataset}.
\newblock UCI Machine Learning Repository, 2014.
\newblock {DOI}: https://doi.org/10.24432/C5GP6D.

\bibitem{banos2014mhealthdroid}
Oresti Banos, Rafael Garcia, Juan~A Holgado-Terriza, Miguel Damas, Hector
  Pomares, Ignacio Rojas, Alejandro Saez, and Claudia Villalonga.
\newblock m{H}ealth{D}roid: a novel framework for agile development of mobile
  health applications.
\newblock In {\em International Work-Conference Ambient Assisted Living and
  Daily Activities (IWAAL)}, pages 91--98, 2014.

\bibitem{barshan2016investigating}
Billur Barshan and Aras Yurtman.
\newblock Investigating inter-subject and inter-activity variations in activity
  recognition using wearable motion sensors.
\newblock {\em The Computer Journal}, 59(9):1345--1362, 2016.

\bibitem{calatrava2023ieeesensors}
Francisco~M. Calatrava-Nicol\'as and Oscar~Martinez Mozos.
\newblock Light residual network for human activity recognition using wearable
  sensor data.
\newblock {\em IEEE Sensors Letters}, 7(10):1--4, 2023.

\bibitem{chaquet2013survey}
Jose~M Chaquet, Enrique~J Carmona, and Antonio Fern{\'a}ndez-Caballero.
\newblock A survey of video datasets for human action and activity recognition.
\newblock {\em Computer Vision and Image Understanding}, 117(6):633--659, 2013.

\bibitem{chen2020metier}
Ling Chen, Yi~Zhang, and Liangying Peng.
\newblock Metier: A deep multi-task learning based activity and user
  recognition model using wearable sensors.
\newblock {\em Proceedings of the ACM on Interactive, Mobile, Wearable and
  Ubiquitous Technologies}, 4(1):1--18, 2020.

\bibitem{Cruciani}
Federico Cruciani, Anastasios Vafeiadis, Chris Nugent, Ian Cleland, Paul
  Mccullagh, Konstantinos Votis, Dimitrios Giakoumis, Dimitrios Tzovaras,
  Liming Chen, and Raouf Hamzaoui.
\newblock Comparing cnn and human crafted features for human activity
  recognition.
\newblock In {\em IEEE SmartWorld/SCALCOM/UIC/ATC/CBDCom/IOP/SCI}, pages
  960--967, 2019.

\bibitem{dang2020sensor}
L~Minh Dang, Kyungbok Min, Hanxiang Wang, Md~Jalil Piran, Cheol~Hee Lee, and
  Hyeonjoon Moon.
\newblock Sensor-based and vision-based human activity recognition: A
  comprehensive survey.
\newblock {\em Pattern Recognit.}, 108:107561, 2020.

\bibitem{debes2016monitoring}
Christian Debes, Andreas Merentitis, Sergey Sukhanov, Maria Niessen, Nikolaos
  Frangiadakis, and Alexander Bauer.
\newblock Monitoring activities of daily living in smart homes: Understanding
  human behavior.
\newblock {\em IEEE Signal Process. Mag.}, 33(2):81--94, 2016.

\bibitem{goodfellow2016nips}
Ian Goodfellow.
\newblock Nips 2016 tutorial: Generative adversarial networks.
\newblock {\em arXiv preprint arXiv:1701.00160}, 2016.

\bibitem{hammerla2015let}
Nils~Y Hammerla and Thomas Pl{\"o}tz.
\newblock Let's (not) stick together: pairwise similarity biases
  cross-validation in activity recognition.
\newblock In {\em Proceedings of the 2015 ACM international joint conference on
  pervasive and ubiquitous computing}, pages 1041--1051, 2015.

\bibitem{Fabio_Hernandez}
Fabio Hernández, Luis~F. Suárez, Javier Villamizar, and Miguel Altuve.
\newblock Human activity recognition on smartphones using a bidirectional lstm
  network.
\newblock In {\em XXII Symposium on image, signal processing and artificial
  vision (STSIVA)}, pages 1--5, 2019.

\bibitem{iwasawa2017privacy}
Yusuke Iwasawa, Kotaro Nakayama, Ikuko Yairi, and Yutaka Matsuo.
\newblock Privacy issues regarding the application of dnns to
  activity-recognition using wearables and its countermeasures by use of
  adversarial training.
\newblock In {\em International Joint Conference on Artificial Intelligence
  (IJCAI)}, pages 1930--1936, 2017.

\bibitem{Khatun}
Mst.~Alema Khatun, Mohammad~Abu Yousuf, Sabbir Ahmed, Md.~Zia Uddin, Salem~A.
  Alyami, Samer Al-Ashhab, Hanan~F. Akhdar, Asaduzzaman Khan, Akm Azad, and
  Mohammad~Ali Moni.
\newblock Deep cnn-lstm with self-attention model for human activity
  recognition using wearable sensor.
\newblock {\em IEEE J. Transl. Eng. Health Med.-JTEHM}, 10:1--16, 2022.

\bibitem{motiian2017few}
Saeid Motiian, Quinn Jones, Seyed Iranmanesh, and Gianfranco Doretto.
\newblock Few-shot adversarial domain adaptation.
\newblock {\em Advances in neural information processing systems}, 30, 2017.

\bibitem{ohn2016looking}
Eshed Ohn-Bar and Mohan~Manubhai Trivedi.
\newblock Looking at humans in the age of self-driving and highly automated
  vehicles.
\newblock {\em IEEE T. Intell. Veh.}, 1(1):90--104, 2016.

\bibitem{ordonez2016deep}
Francisco~Javier Ord{\'o}{\~n}ez and Daniel Roggen.
\newblock Deep convolutional and lstm recurrent neural networks for multimodal
  wearable activity recognition.
\newblock {\em Sensors}, 16(1):115, 2016.

\bibitem{scikit-learn}
F.~Pedregosa, G.~Varoquaux, A.~Gramfort, V.~Michel, B.~Thirion, O.~Grisel,
  M.~Blondel, P.~Prettenhofer, R.~Weiss, V.~Dubourg, J.~Vanderplas, A.~Passos,
  D.~Cournapeau, M.~Brucher, M.~Perrot, and E.~Duchesnay.
\newblock Scikit-learn: Machine learning in {P}ython.
\newblock {\em Journal of Machine Learning Research}, 12:2825--2830, 2011.

\bibitem{peng2018aroma}
Liangying Peng, Ling Chen, Zhenan Ye, and Yi~Zhang.
\newblock Aroma: A deep multi-task learning based simple and complex human
  activity recognition method using wearable sensors.
\newblock {\em Proceedings of the ACM on Interactive, Mobile, Wearable and
  Ubiquitous Technologies}, 2(2):1--16, 2018.

\bibitem{pirsiavash2012detecting}
Hamed Pirsiavash and Deva Ramanan.
\newblock Detecting activities of daily living in first-person camera views.
\newblock In {\em IEEE/CVF Computer Vision and Pattern Recognition Conference
  (CVPR)}, pages 2847--2854, 2012.

\bibitem{piyathilaka2015human}
Lasitha Piyathilaka and Sarath Kodagoda.
\newblock Human activity recognition for domestic robots.
\newblock In {\em Field and Service Robotics: Results of the 9th International
  Conference}, pages 395--408. Springer, 2015.

\bibitem{Plotz2023Percom}
Thomas Plötz.
\newblock If only we had more data!: Sensor-based human activity recognition in
  challenging scenarios.
\newblock In {\em IEEE International Conference on Pervasive Computing and
  Communications (PerCom) Workshops}, pages 565--570, 2023.

\bibitem{qin2023generalizable}
Xin Qin, Jindong Wang, Shuo Ma, Wang Lu, Yongchun Zhu, Xing Xie, and Yiqiang
  Chen.
\newblock Generalizable low-resource activity recognition with diverse and
  discriminative representation learning.
\newblock In {\em Proceedings of the 29th ACM SIGKDD Conference on Knowledge
  Discovery and Data Mining}, pages 1943--1953, 2023.

\bibitem{rasouli2019autonomous}
Amir Rasouli and John~K Tsotsos.
\newblock Autonomous vehicles that interact with pedestrians: A survey of
  theory and practice.
\newblock {\em IEEE Trans. Intell. Transp. Syst.}, 21(3):900--918, 2019.

\bibitem{Reiss2012}
Attila Reiss and Didier Stricker.
\newblock Introducing a new benchmarked dataset for activity monitoring.
\newblock In {\em International Symposium on Wearable Computers}, pages
  108--109, 2012.

\bibitem{roitberg2014human}
Alina Roitberg, Alexander Perzylo, Nikhil Somani, Manuel Giuliani, Markus
  Rickert, and Alois Knoll.
\newblock Human activity recognition in the context of industrial human-robot
  interaction.
\newblock In {\em Signal and Information Processing Association Annual Summit
  and Conference (APSIPA)}, pages 1--10. IEEE, 2014.

\bibitem{ShuSungho2022Percom}
Sungho Suh, Vitor~Fortes Rey, and Paul Lukowicz.
\newblock Adversarial deep feature extraction network for user independent
  human activity recognition.
\newblock In {\em IEEE International Conference on Pervasive Computing and
  Communications (PerCom)}, pages 217--226, 2022.

\bibitem{Ullah}
Mohib Ullah, Habib Ullah, Sultan~Daud Khan, and Faouzi~Alaya Cheikh.
\newblock Stacked lstm network for human activity recognition using smartphone
  data.
\newblock In {\em European workshop on visual information processing (EUVIP)},
  pages 175--180, 2019.

\bibitem{wang2019survey}
Yan Wang, Shuang Cang, and Hongnian Yu.
\newblock A survey on wearable sensor modality centred human activity
  recognition in health care.
\newblock {\em Expert Syst. Appl.}, 137:167--190, 2019.

\bibitem{YangWang}
Yang Wang, Hongji Xu, Yunxia Liu, Mengmeng Wang, Yuhao Wang, Yang Yang, Shuang
  Zhou, Jiaqi Zeng, Jie Xu, Shijie Li, and Jianjun Li.
\newblock A novel deep multi-feature extraction framework based on attention
  mechanism using wearable sensor data for human activity recognition.
\newblock {\em IEEE Sens. J.}, 23(7):7188--7198, 2023.

\bibitem{yang2015deep}
Jianbo Yang, Minh~Nhut Nguyen, Phyo~Phyo San, Xiaoli Li, and Shonali
  Krishnaswamy.
\newblock Deep convolutional neural networks on multichannel time series for
  human activity recognition.
\newblock In {\em International Joint Conference on Artificial Intelligence
  (IJCAI)}, volume~15, pages 3995--4001. Buenos Aires, Argentina, 2015.

\bibitem{yeffet2009local}
Lahav Yeffet and Lior Wolf.
\newblock Local trinary patterns for human action recognition.
\newblock In {\em IEEE 12th international conference on computer vision
  (ICCV)}, pages 492--497, 2009.

\bibitem{zhou2020deep}
Xiaokang Zhou, Wei Liang, I~Kevin, Kai Wang, Hao Wang, Laurence~T Yang, and Qun
  Jin.
\newblock Deep-learning-enhanced human activity recognition for internet of
  healthcare things.
\newblock {\em IEEE Internet Things J.}, 7(7):6429--6438, 2020.

\end{thebibliography}

\end{document}